\documentclass[12pt]{article}
\pdfoutput=1
\usepackage{graphicx}
\usepackage{psfrag}
\usepackage{enumerate}
\usepackage{soul}
\usepackage{subfig}
\usepackage{jheppub}

\newcommand{\be}{\begin{equation}}
\newcommand{\ee}{\end{equation}}
\newcommand{\bea}{\begin{eqnarray}}
\newcommand{\eea}{\end{eqnarray}}

\def\[{\left [}
\def\]{\right ]}
\def\({\left (}
\def\){\right )}

\def\r2{\sqrt{2}}

\newcommand{\bbibitem}[1]{\bibitem{#1}\marginpar{#1}}

\def\Label#1{\label{#1}%
  \smash{\hbox to0pt{\raise1ex\hbox{\tiny[#1]}\hss}}}
\def\noLabels{\let\Label=\label}
\def\nobbibitem{\let\bbibitem=\bibitem}

\begin{document}

\noLabels
\nobbibitem

\DeclareGraphicsExtensions{.pdf,.png,.gif,.jpg,.eps}

\begin{titlepage}

\begin{center}
{\Large \bf Flux Discharge Cascades in Various Dimensions}
\vspace{6mm}

{Matthew Kleban, Karthik Krishnaiyengar and Massimo Porrati}

\vspace{6mm}
{\it Center for Cosmology and Particle Physics\\Department of Physics, 
New York University \\
New York, NY 10003, USA}

\end{center}
%------------------------------------------------------------------------------

%\setcounter{footnote}{0}
%%%%%%%%%%%%%%%%%%%%%%%%%%%%%%%%%%%%%%%%%%%%%%%%%%%%%%%%%%%%%%%%%%%%%%%%%%%%%%%%%%%%%%%
\begin{abstract}
\noindent
We study the dynamics of electric flux discharge by charged particle pair or spherical string or membrane production in various dimensions.  When electric flux wraps at least one compact cycle, we find that a single ``pair" production event can initiate a cascading decay in real time that ``shorts out" the flux and discharges many units of it.  This process arises from local dynamics in the compact space, and so is invisible in the dimensionally-reduced truncation.  It occurs in theories as simple as the Schwinger model on a circle, and has implications for any theory with compact dimensions and electric flux, including string theories and the string landscape.
\end{abstract}
%%%%%%%%%%%%%%%%%%%%%%%%%%%%%%%%%%%%%%%%%%%%%%%%%%%%%%%%%%%%%%%%%%%%%%%%%%%%%%%%%%%%%%%%%
%\vspace{0.5in}
\end{titlepage}

%\renewcommand{\baselinestretch}{1.05}  %Line spacing
%%%%%%%%%%%%%%%%%%%%%%%%%%%%%%%%%%%%%%%%%%%%%%%%%%%%%%%%%%%%%%%%%%%%%%%%%%%%%%%%
%%%%%%%%%%%%%%%%%%%%%%%%%%%%%%%%%%%%%%%%%%%%%%%%%%%%%%%%%%%%%%%%%%%%%%%%%%%%%%%%%%%%%%%%%%%
\tableofcontents %uncomment to display table of contents

\section{Introduction}

It is well known that electric flux can decay via the Schwinger process: the non-perturbative nucleation of charged particle-anti particle pairs \cite{Schwinger:1951nm}.  This phenomenon occurs in theories with any number of spatial dimensions, and has a natural generalization to theories with higher form gauge theories, where the decay proceeds by the nucleation of spherical p-branes of zero net charge \cite{Brown:1988kg}.  

Perhaps the best studied model in this class is the massive Schwinger model---QED in 1+1 dimensions with massive electrons \cite{Schwinger:1962tp,Coleman:1975pw, Coleman:1976uz}.  Most of the very extensive body of past work has focussed on its spectrum and other static properties, while comparatively little attention has been paid to dynamics.  As we will see, when the spatial direction of the Schwinger model is compact, the dynamics of electric flux decay has a surprising twist that generalizes in an interesting way to higher dimensional models.

One of the charateristic features of the Schwinger model is that electric flux in 1+1 dimensions---that is, a constant field strength $F_{01}$, which satisfies the equations of motion for the theory in the absence of any charges---acts like a cosmological constant.  The reason is physically obvious:  there are no electromagnetic excitations in 1+1 dimensions because there are no transverse directions for them to be polarized in; moreover, there is no ``room'' for flux lines to spread, and therefore the energy density $\rho_{F}=F^{2}/2$ is constant and remains so when the space expands---it is a vacuum energy.

In general, the same is true of any top-form ($d$-form in $d$-spacetime dimensions) field strength---it has no local dynamics, and a non-zero background value (which is necessarily ``electric", since one index on any non-zero element must be time) acts as a vacuum energy.  Lower form electric fluxes can also act like cosmological constants when they have ``legs'' in all the non-compact spatial dimensions and the compact dimensions are stabilized.  This is the origin of  vacuum energy in {\it e.g.} the Freund-Rubin solutions \cite{Freund:1980xh},  the toy string landscape model of Bousso and Polchinski \cite{discretuum}, as well as more realistic string vacua such as \cite{kklt}.

In this note we describe a new mechanism in the dynamics of the decay of a $N$ units of initial-state  electric flux in the massive Schwinger model compactified on a spatial circle, and in higher dimensional models with $p$-form electric flux. 
We consider the semi-classical limit in which the electron-positron pair or charged brane nucleation rate is slow compared to other relevant time scales, so that we can focus on the time evolution of a single nucleation event.  

We find that when there are compact spatial dimensions,  a single nucleation event that initially discharges only a single unit of flux in a small region can initiate a cascade that results in the discharge of all  the flux, and in some cases can even lead to an oscillating solution where the flux varies between $N$ and $-N$.    In the Schwinger model we will study this in both the Dirac representation (in which the degree of freedom is a  massive fermion coupled to the electric field) and the massive sine-Gordon representation (in which the degree of freedom is a real scalar field with a sinusoidal potential).  Because it relies on local dynamics as the field intersects itself, the cascade can occur even when the potential energy of the false vacuum the field tunnels from is well below the energy of the potential barrier containing the vacuum the field tunnels to.

We will study models with gravity switched off---in other words, we will ignore the gravitational backreaction due to the energy in the charged particles/branes and due to the electric flux.  In many models, the energy in the flux makes an important contribution to the geometry, for instance stabilizing the shape or size of some extra dimensions or as a significant source of 4D vacuum energy.  In these models our flux cascade would be strongly modified by the inclusion of backreaction, at least once it has proceeded to the point that a significant fraction of the flux has been discharged.  We will leave the implications of that for future work.

\paragraph{Relation to previous work:} Pair creation of strings and branes in electric fields was studied in \cite{Brown:1988kg, Bachas:1992bh}.  

Past work has shown that in some cases instantons exist that discharge multiple units of flux \cite{Brown:2010bc,Brown:2010mg}.  These transitions typically occur in models where there are a few units of many different types of flux.  Instead, the mechanism we will discuss occurs when there are many units of a single type of flux, and in models as simple as the Schwinger model on a circle.  Another important difference is that our mechanism is dynamical---it takes place in real time after the nucleation event mediated by the instanton (which itself discharges only a single unit of flux).  Roughly speaking, the mechanism of  \cite{Brown:2010bc} is important when the charged objects are heavy enough that gravitational self-interactions are important, whereas our mechanism occurs when they are light enough that the size of the instanton is smaller than at least some compact extra dimensions.  It would be interesting to explore how much parameters space remains between these two limits.

In studying the dynamicals of bubbles of reduced flux, we will use a mechanism that is related to one first studied in \cite{BlancoPillado:2009di, Easther:2009ft, Giblin:2010bd}.  The idea of bubbles colliding with themselves after wrapping a compact dimension was explored in \cite{Brown:2008ea} as a potential solution to the problems of old inflation.

\section{Instantons in the Massive Schwinger Model}

The massive Schwinger model \cite{Schwinger:1962tp} is quantum electrodynamics with massive electrons in 1+1 dimensions.  It is parametrized by a mass $m$ and a charge $e$, both with dimensions of mass, as well as a theta angle that we will set to zero.  There are two  ways to represent the model: the standard QED-like representation with a $U(1)$ gauge field and a massive Dirac fermion field, and a bosonized action with a single self-interacting scalar degree of freedom---the massive sine-Gordon model \cite{Coleman:1975pw, Coleman:1976uz}. 

For notational simplicity we will assume that the charge $e>0$, and when we consider a background field, $E>0$.

\subsection{Dirac Representation}

Because there are no transverse dimensions in which flux can spread, classically a single charged particle in 1+1 dimensions creates a {\it constant} electric field:
\be
E(x) = e~ {\rm sgn}(x),
\ee
where $e$ is the charge and the particle is located at $x=0$.  As a result,  two charged particles exert forces on each other that are independent of the distance between them.  For closely related reasons, there are no magnetic fields ($F_{\mu \nu} = E \epsilon_{\mu \nu}$) and the gauge constraint removes all propagating on and off-shell degrees of freedom from the gauge field---there are no electromagnetic waves or photons.  This makes the quantum Schwinger model, defined by the Lagrangian density  (with $\theta=0$)
\be \label{schwinger}
{\cal L} = -\bar{\psi}(i\partial_{\mu}\gamma^{\mu} + eA_{\mu}\gamma^{\mu} + im)\psi + \frac{1}{4} F_{\mu\nu}F^{\mu\nu},
\ee
very simple to analyze.  

We will be interested in the regime where all relevant length scales are significantly longer than the Compton wavelength $1/m$ of the electron, and where $e^{+}e^{-}$ pair creation is a rare event that can be controlled semi-classically.  Specifically, this means that
\be \label{approx}
m \gg e, E
\ee
where $e$ is the electron charge and $E$ is the background field strength.  In this limit we can approximate \eqref{schwinger} with the action for a fixed number of charged particles interacting through the electric field.  
For a single particle, the action is  
\be \label{pp}
 S = -m \int \sqrt{-\dot{z}^{\mu } \dot{z_{\mu }}} \, ds + e \int A_{\mu } \dot{z}^{\mu } \, ds - \frac{1}{4} \int F_{\mu \nu } F^{\mu \nu } \, d^2x + \int \partial_\mu \left( A_{\nu } F^{\mu \nu } \right) \, d^2x
 \ee
where ${z^{\mu }}(s)$ is the spacetime trajectory of the  particle as a function of a proper time $s$ (the generalization to multiple particles is obvious).   The last term is \eqref{pp} is necessary for consistency of the variational principle in the presence of a charge; see \cite{Brown:1988kg} for more details.

The equations of motion that follow from \eqref{pp} are \cite{Brown:1988kg}
\bea \label{eom}
 {\ddot  z}^{\mu } = \frac{e}{m} E_{on} \epsilon^{\mu \nu } \dot{z}_{\nu } \\
 \partial_{\mu}F^{\mu\nu} = - e \int \delta^2(x-z(s)) \dot{z}^{\nu} \,ds. 
 \eea
The second equation implies that the electric field is constant everywhere except when crossing a charged particle world line, where it jumps by $e$.  The electric field {\it on} the particle worldline $E_{on}= E(z(s))$ can be defined as the average of the fields on either side of the world line: $E_{on}=(E_{r}+E_{l})/2$, where $E_{r}(E_{l})$ is the electric field just to the right (left) of the particle.  

The solution to the first equation is simply 
\bea \label{hyp} 
z^0 (s) -z^{0}(0)=  \left( \frac{e}{m} E_{on} \right)^{-1} \sinh(\frac{e}{m} E_{on} s)  \\
  z^1 (s) -z^{1}(0) =  \left(\frac{e}{m} E_{on}\right)^{-1} \cosh(\frac{e}{m} E_{on} s) 
  \eea
where again $E_{on}=(E_{r}+E_{l})/2$ and the second equation of motion requires that $E_{r}-E_{l}=e$ (and an analogous solution exists for anti-particles).  Therefore a single charged particle in a non-zero background field ($E_{on}\neq 0$) accelerates with constant proper acceleration  $a= e E_{on}/m$.  If other charges are present, the field $E_{on}$ (and therefore the acceleration of a given charge) can change when their worldlines cross.

\subsection{Instantons in the Dirac Representation} \label{Dirinst}

In a background electric field, $e^{+}e^{-}$ quantum pair creation will occur if the process conserves the total energy.  In the Schwinger model this is not guaranteed: if the magnitude of the background field $|E|<e/2$, due to \eqref{eom} the field in between the two particles after they appear would necessarily be larger in magnitude than the background field.  The energy in the field in between the pair would be higher than it was before the nucleation, and therefore no energy is available to account for the rest mass energy $2m$ of the  particles.  Therefore, the pair creation rate is zero for  $|E|<e/2$.

This constraint,  the rate for the nucleation, and the subsequent Lorentzian evolution can all be derived easily starting from the action \eqref{pp}.  Following the standard procedure, one looks for instanton solutions by analytically continuing the action to Euclidean space \cite{Coleman:1978ae}.  Starting from the solution \eqref{hyp}, it is apparent that the Euclidean-signature solution for the particle worldline is simply a circle with radius 
\be \label{Dircrit}
R_{c}={m \over e E_{on}} = {m \over e (E-e/2)},
\ee
which is therefore an instanton.  

As a check, a simple computation gives the Euclidean action for a circular worldline of radius $R$ in the presence of a background field $E$ \cite{Brown:1988kg}:
\be
S_E = m \int \sqrt{-\dot{z}^{\mu } \dot{z_{\mu }}} \, ds  +  \frac{1}{4} \int F_{\mu \nu } F^{\mu \nu } \, d^2x = \\
2 \pi m R +  \pi R^{2} \left( E-e \right)^{2}/2 + \pi E^{2} \int_{R}^{\infty}  r dr
\ee
The probability to nucleate a pair is $P(pair) \propto e^{-B}$, where $B(R) = (S_E({\rm instanton}) - S_E({\rm background}))$.  Subtracting the background gives
$$
B(R)=2 \pi m R - \pi R^{2} e(E-e/2) = \pi R(m-e E_{on} R). $$
Minimizing with respect to $R$ gives $R_{c}=m/(e E_{on})$ [in agreement with \eqref{Dircrit}], and 
\be \label{action}
B={ \pi m^{2} \over e (E-e/2)}.
\ee
As promised, in the limit \eqref{approx} $B \gg 1$, so pair production is exponentially rare.  It is simple to check that the pair-production event conserves energy.

\subsection{Bose Representation}

The Lagrangian \eqref{schwinger} is equivalent to the theory of a single massive scalar in a cosine potential:
\be \label{msg}
{\cal L} = \frac{1}{2} (\partial_{\mu}\phi)^2 + \frac{e^2}{2\pi}\phi^2 - c~m^2 \cos{(2\sqrt{\pi}\phi)},
\ee
where $\phi$ is a real scalar and $c$ is an ${\cal O}(1)$ constant \cite{Coleman:1974bu,Coleman:1975pw}.
The relation to the Dirac form of the theory is
$$:\bar{\psi}\psi: = -c m \cos{(2\sqrt{\pi}\phi)}$$
$$j^{\mu} = :\bar{\psi}\gamma^{\mu}\psi: = \pi^{-1/2}\epsilon^{\mu\nu}\partial_{\nu}\phi$$
$$F_{01} = e \pi^{-1/2}\phi.$$
The equation for the current $j$ shows that spatial ``kinks''  (``anti-kinks'') in the scalar correspond to positive (negative) charges.  

\begin{figure}
\subfloat[][]{\label{subfig:my-a}\includegraphics[width=.45\textwidth]{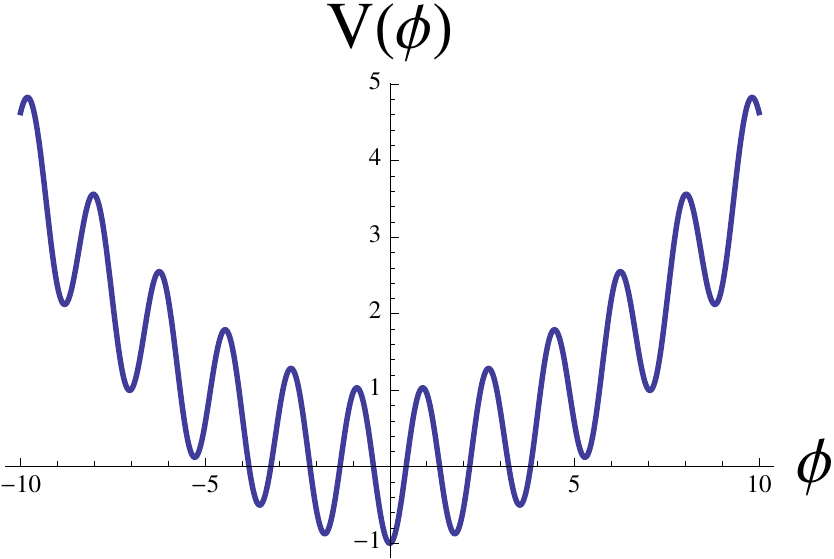}} \hspace{.5 in}
\subfloat[][]{\label{subfig:my-b}\includegraphics[width=.45\textwidth]{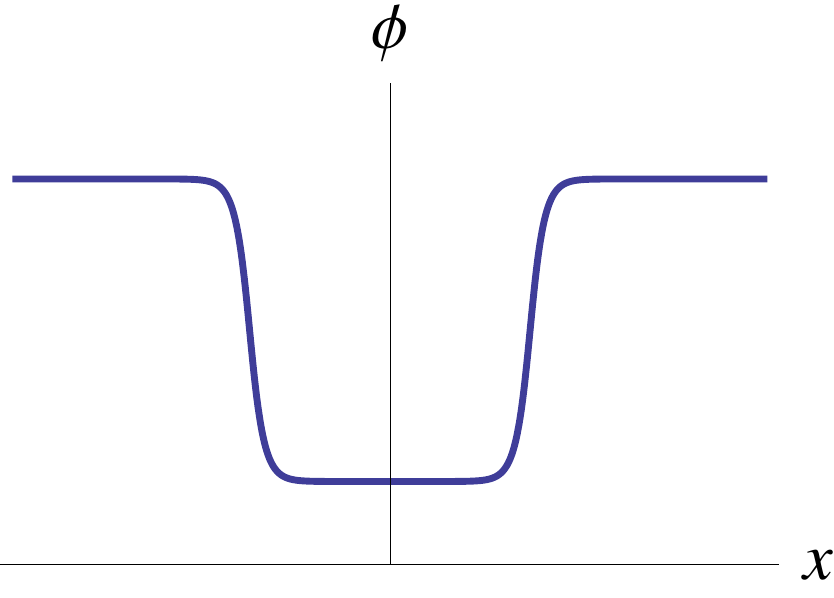}} 
\caption{ \subref{subfig:my-a} The potential for the massive sine-Gordon model. 
\subref{subfig:my-b} A kink-anti-kink pair in the presence of background flux.}
\end{figure}

\subsection{Sine-Gordon Instantons}
Because of the quadratic term $e^{2} \phi^{2}/(2 \pi)$, the  minima $\phi_{n}\approx n \sqrt{\pi}$  of the potential $V(\phi)= \frac{e^2}{2\pi}\phi^2 - c~m^2 \cos{(2\sqrt{\pi}\phi)}$ have $n$-dependent energies.  As such, with the exception of the true minimum $\phi(x)=\phi_{0}=0$, spatially homogeneous configurations of $\phi$ are at best metastable.  The tunneling of the scalar field from one minima to another can be understood with standard instanton techniques \cite{Coleman:1978ae}.  The simplest scenario is for the sine-Gordon model on an infinite line, and when the thin wall limit for the instantons is valid (which simply requires $e/m \ll 1$, as we wil see).  

In the thin wall limit, the Euclidean instanton that mediates tunneling from  $\phi=\phi_{n}$ to $\phi_{n-1}$ is a disk of radius $R_{c}$  bounded by a thin wall.   For $r<R_{c}, \phi \approx \phi_{n-1}$, and for $r>R_{c}, \phi \approx \phi_{n}$.  A radial slice through the wall is a ``kink'' of the sine-Gordon model; that is, it is an electron or positron (depending on the direction of the slice), and the decrease in $\langle \phi \rangle$ inside the disk corresponds to the change in the electric field $E$ of the Dirac representation.

In the thin wall approximation, we can find the critical radius $R_{c}$ of the bubble by minimizing the Euclidean action 
$$S_{E} = - \pi R^{2} \epsilon + 2 \pi R \sigma,$$
where $\epsilon = V(\phi_{n})-V(\phi_{n-1})$ is the energy difference between the two minima, $\phi_{n} \equiv n \sqrt{\pi}$, and $\sigma =  \int_{\phi_{n-1}}^{\phi_{n}} d\phi \sqrt{2V(\phi)-2V(\phi_{n-1})}$ is the surface tension.\footnote{In the sine-Gordon model near the free-fermion point, there is an ${\cal O}(1)$ quantum correction to the mass of the kink.  However, this can simply be re-absorbed into the definition of $c$.  All other quantum corrections to the instanton solution should be suppressed by powers of $(e/m)^{2}$.}  In the limit $m \gg e$ and for an appropriate choice of $c$, one obtains $\epsilon = e^{2}(n-1/2)$ and $\sigma = m$, which gives
\be
R_{c}={m \over e^{2}(n-1/2)} = {m \over e(E-e/2)},
\ee 
in agreement with \eqref{Dircrit}.  The thickness of the wall is $\sim 1/m$, while in the limit \eqref{approx} the bubble radius $R_{c} >> 1/m$, justifying the thin wall approximation.  The action is  simple to compute and agrees with \eqref{action}.

At very large $\phi$, the minima disappear due to the effect of the quadratic term.  This happens when $dV/d\phi=0$ can no longer be solved; namely when
\be
 \phi \sim (m/e)^2  \leftrightarrow E \sim m^2/e.
\ee
  Not surprisingly, this is the value of $E$ where the instanton action \eqref{action} becomes ${\cal O}(1)$.

\subsection{Instantons on a Circle}

We are interested in the Schwinger model on a spatial circle of radius $L$.  So long as $R_{c} \ll L$, the instanton solutions described in the previous sections should be approximately unaffected by the boundary conditions.  
As one decreases $L$, the circular instantons will deform into an ellipsoidal shape and eventually merge with their covering space neighbors.  For very small $L$, in the bosonized form of the theory \eqref{msg} one can see easily that the instanton becomes independent of the spatial coordinate $x$: it consists of two kinks in Euclidean time direction $\tau$, separated by a critical distance of order $R_{c}$  (a similar phenomenon occurs in instantons  at finite temperature, with the difference being that in that case the compact direction is Euclidean time; see \cite{Linde:1980tt}).

Therefore in the limit of small $L$ the transition does not proceed via the nucleation of a localized $e^{+}e^{-}$ pair or bubble, but instead by an event which changes the field configuration homogeneously; everywhere on the circle at one instant.  Because the resulting field is independent of position, the dynamics are simply those of a particle in the potential $V(\phi)$.  If the initial state field is $\phi_{n} \approx n \sqrt{\pi}$, the transition produces a field configuration in which $\phi$ has tunneled to the opposite side of the barrier to the point with energy equal to its starting point, {\it i.e.} after the tunneling  $\phi = \phi_{n-1} + \epsilon$ everywhere on the circle, where $V(\phi_{n-1} + \epsilon) = V(\phi_{n})$.  What happens next depends on the height of the barrier around $\phi_{n-1}$.  If the barrier height is higher than $V(\phi_{n})$, the subsequent time evolution will be periodic oscillations around $\phi=\phi_{n-1}$ and there is no flux discharge cascade.   This state is similar to a breather mode of the sine-Gordon model, with a wavefunction spread uniformly around the circle. 

If the  height of the barrier separating $\phi_{n-1}$ and $\phi_{n-2}$ is below $V(\phi_{n})$, there is enough energy for the field to simply flow over the maximum and continue down.  This leads to a different sort of flux discharge cascade, one which exists even in the dimensionally reduced theory.  However this happens when $e^{2} (\phi_{n}^{2}-\phi_{n-1}^{2})/2\pi \approx cm^{2}$, which means
\be
  \phi \approx  (m/e)^{2}
 \ee
and the action for the instanton is ${\cal O}(1)$ and the semi-classical analysis cannot be trusted in any case.

\section{Lorentzian Evolution and Flux Cascades}

Consider the massive Schwinger model with an initial-state background field $E_{N} = Ne$ on a spatial circle of length $L$, with $R_{c}=m/[e(E_{N}-e/2)] \ll L$ so that the $E$-field can decay by localized $e^{+}e^{-}$ nucleation.  The initial state immediately after such a nucleation is determined by a slice of the Euclidean instanton solution: the particles appear at rest a distance $2 R_{c}$ apart, and then begin to accelerate away from each other due to the force exerted by the remaining electric field $E_{on}=e(N-1/2)$.  From the solution \eqref{hyp}, one sees that after a time $t \sim L/2$ [$s \sim (eE_{N}/m)\ln eE_{N}L/2 m$] the particles will collide at the point on the circle opposite where they nucleated.  The collision is relativistic in the limit $R_{c} \ll L$:
\be
\beta={dx \over dt} = \tanh {m s \over e E_{N}} \approx \tanh \ln {eE-NL \over 2m} \approx 1 - 2 \left( {2m \over e^{2}N L} \right)^{2},
\ee
so 
\be
\gamma =  {e E_{N} L \over 4 m} = {e^{2} N L  \over 4 m} \approx {L \over 4 R_{c}}
\ee
where the approxmations are valid for $N \gg 1$, $m \gg Ne$, and $L \gg R_{c}$. 
%CHECK FACTOR OF 4	

When the electron and positron collide, {\it a priori} a number of different things could happen.  One is annihilation into some other form of energy.  Another is reflection, and a third is free passage---the particles transmit through each other more or less unaffected.  Let us examine these possibilities in turn.

The massive Schwinger model has an extremely limited set of excitations.  Indeed, in the pure sine-Gordon model on a line the spectrum is known exactly, as is the S-matrix element for kink-anti-kink scattering.  For a  kink-anti-kink collision, due to  conservation laws there is zero probability for any final state other than kink-anti-kink, and so (in the massless sine-Gordon model) annihilation or radiation is impossible.  As for reflection versus transmission, an exact $S$-matrix result shows that the probability for transmission divided by the probability for reflection grows quadratically with the center of mass energy: $P(T)/P(R)\sim s^2 $ for $s \gg m^2$ \cite{Zamolodchikov:1977py}.  
 This conclusion (that reflection is unlikely) is quite natural from the point of view of the classical Dirac theory, since the force between charged particles is independent of their separation.  It also has a natural interpretation in the classical sine-Gordon model, as we will discuss later.

The massive Schwinger model on a circle is not quite as simple, due to the mass term---proportional to $e^{2}$---in the bosonized representation, and to the periodic boundary conditions.  Nevertheless, in the limit under consideration, where $e \ll m$ and $R_{c} \ll L$, we expect the same conclusions to hold---that with high probability the particles will simply pass through each other.\footnote{Even if they reflect, the acceleration due to the $E$ field will bring them back to another collision after a time of order $L$.}  Annihilation back to the initial state of homogeneous electric field $E$, while possible, should have very small amplitude: it would require a change in the field everywhere on the circle (including far from the collision), and there is no instanton that mediates such a transition.  Annihilation to a state with homogeneous field $E-e$ is impossible as it does not conserve energy.  Transmission or reflection plus production of some number of other particles is possible.  The probability for such production must be suppressed by a power of $(e/m)^{2}$, which is small in the limit we are considering.  However in the strong coupling limit $e/m \sim 1$ such effects can be important, and could certainly alter our conclusions.  As we will discuss, such considerations apply to higher dimensional theories as well.

If the particles do pass through one another, what is the state of the electric field in between them afterwards?  It is very simple to see that it can only be $E_{N-2}=(N-2)e=\phi_{N-2}$.  In the Dirac representation this follows from Gauss' law and locality.  The field on the far side of the circle at the time of the collision was $E_{N}-e=(N-1)e=E_{N-1}$.  By locality, the field far away does not change at the time of the collision.  On the other hand, Gauss' law \eqref{eom} guarantees that the field must jump by $e$ across each particle.  Therefore, the field in between the $e^{+}$ and $e^{-}$ after they have passed through each other must be $E_{N}-2e$:  the field has been discharged by two units.  So long as $N>3/2$, eq.~\eqref{eom} shows that there will still be a force accelerating the particles in the same direction as before.  Therefore they will collide again at the antipodal point on the circle, in which case the same analysis applies, but with $N \rightarrow N-1$ and even larger kinetic energy for the particles.  Iterating the above argument shows that all $N$ units of flux will discharge---and that the process does not end there. 

\begin{figure} \label{cicle}
\hspace{0.5 in}\includegraphics[width=.8\textwidth]{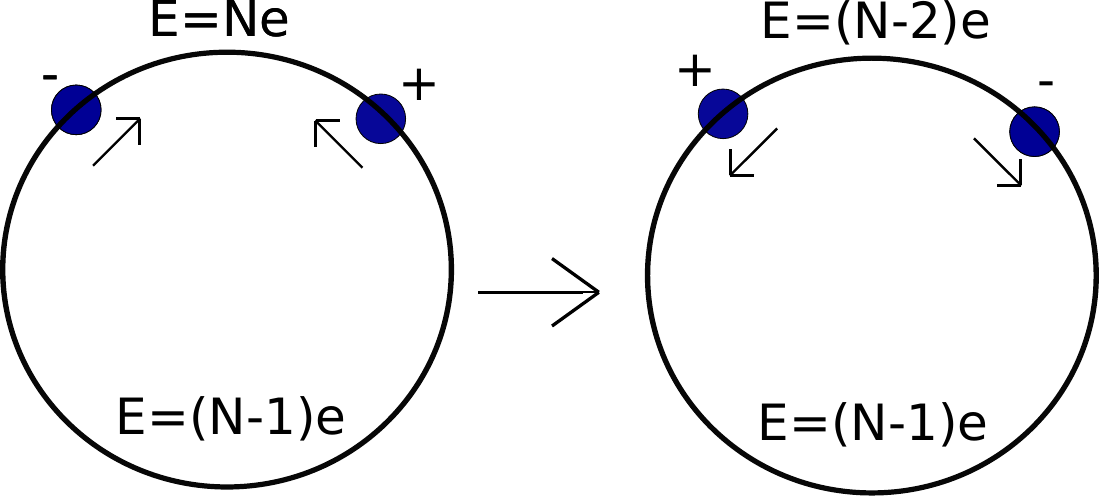}
\caption{\label{fig-temp}Figure showing the evolution of an electron-positron pair in the Schwinger model on a circle, in the presence of a background field $E=Ne$.  If the particles transmit past each other, the flux in between them is discharged by $2e$ relative to the initial field.}
\end{figure}

A picturesque way to visualize the flux cascade is to think of the electric flux as an elastic band initially wrapped $N$ times around the circle.  Pair creation cuts out a length $2 R_{c}$ of the band, and stores its energy in two massive endpoints (the $e^{+}e^{-}$ pair).  The tension in the band accelerates the ends around the circle, unwinding it until all $N$ winds have unwrapped and all the energy is entirely kinetic, stored in the momentum of the ends.  

This   makes it easy to see what happens next.  The kinetic energy in the particles is very large---and as per the argument above, the particles are therefore very likely to transmit through each other on their next collision.  If so, since they trail flux behind them they will begin to { \it increase} the field strength, but now with the opposite sign as in the initial state.  Conservation of energy dictates that, absent some other interaction (like a reflection, emission of radiation, or back-reaction due to gravitational effects), the field will ramp back up until it reaches the negative of its initial value $E_{-N} = -E_{N}=-N e$---at which point the particles are instantaneously at rest a distance $2 R_{c}$ apart, and then the whole process repeats.

In this manner the field as a whole can oscillate back and forth between $E=E_{N}$ and $E=-E_{N}$, with each oscillation taking a time of order $N L$.  So long as the decay time $e^{+B}/m\sim e^{m^{2}/(N e^{2})}/m \gg N L$, many such oscillations will take place.  On exponentially long time scales $\sim e^{+B}$, additional pair creation events will complicate the situation, and perturbative radiation or particle creation may add a ``friction'' term that ends the oscillation.  

At first glance in the bosonized form this looks like the zero mode of the field $\phi$ oscillating in the harmonic part of its potential $e^{2} \phi^{2}$ as if ignoring the ``bumps''---but the period is $NL$, not $1/e$, and indeed one cannot ignore the dynamics on the spatial circle.

\subsection{Free Passage}

Another argument for the near-perfect transmission of the scalar field kink-anti-kink pairs comes from classical field theory \cite{Giblin:2010bd}.  When the kinetic energy in two co-dimension one scalar field walls is large compared to any relevant potential energy barriers, one should be able to ignore the potential energy and treat the field as free.  Consider a configuration of two colliding free-field codimension-1 walls.  In between the walls before the collision the field takes the value $\phi=\phi_{F}$, while behind the walls to the left (right) the field takes the values $\phi = \phi_{L}$  ($\phi=\phi_{R}$).  A simple argument shows that after the collision, the field in the overlap region between the walls immediately after the collision will be $\phi = \phi_{F}+(\phi_{L}-\phi_{F}) + (\phi_{R}-\phi_{F})$ (this equation also applies as a vector equation for multiple scalar fields).  As  time passes, the evolution of the field in the central region will be determined by its potential energy functional (and  any other interactions the field may be subject to).  In Appendix \ref{AA}, we generalize this argument to the case of spherically symmetric collapsing regions in arbitrary spacetime dimension.

In our case, at each collision $\phi_{F} = \phi_{n} \approx n \sqrt{\pi}$, and $\phi_{L} = \phi_{R} \approx (n-1)\sqrt{\pi}$.  Therefore, the field in the overlap region should be close to $(n-1-1)\sqrt{\pi}=(n-2)\sqrt{\pi} \approx \phi_{n-2}$.  Since (at least for $n>1$) $\phi_{n-2}$ is a minimum with potential energy lower than $\phi_{n-1}$, the expansion of this region will accelerate due to the outward force on its walls.

\section{Higher Dimensional Models}

This flux cascade process has many generalizations to  higher dimensional theories.  A necessary condition for it to occur is the presence of electric flux wrapped on at least one compact spatial dimension (i.e., a field strength with one ``leg'' in the time direction and at least one more in a compact spatial direction).  This is not a sufficient condition $R_c > L$ limit in the Schwinger model illustrates.  Indeed, if the compact direction is small one should be able to dimensionally reduce on it, in which case dynamics that depend on motion in the compact direction should be irrelevant.  Hence, another condition is that the size of the instanton be smaller than the size of the compact dimension(s), $R_{c} \ll L$.

Even in cases where the bubble collides with itself, interactions might prevent it from passing through itself, or it could dissipate so much energy into other degrees of freedom that it cannot climb over the barrier and discharge more flux.  These considerations are strongly model-dependent, so we will not attempt to treat them in general.  Instead, we will present a series of examples that are illustrative of the various phenomena that can occur. 

Our essential point is that once a bubble appears, the flux cascade process always occurs in the free limit where the bubble walls do not interact.  Therefore in any model where interactions between the walls are weak, one expects that they are unlikely to prevent it from happening.  In very general terms, theories with flux are metastable, and once a significant perturbation appears---like a bubble inside of which one unit of flux is discharged---it is not particularly surprising that its dynamics can stimulate additional decays to a lower energy state.

Under some circumstances magnetic flux can also discharge in a cascade.  Here, by ``magnetic'' flux we mean a background field strength satisfying $F^{2}<0$.  If the theory contains magnetic monopoles or branes ({\it i.e.} objects that act as sources for the magnetic flux \cite{Affleck:1981ag}), and if the dualized flux (which is necessarily electric in the sense that $\tilde F^{2} > 0$, where $\tilde F \equiv *F$) satisfies the conditions worked out here, then the cascade should occur.

\subsection{$2+1$ Dimensional Models}

\paragraph{Cylinder with 3-Form Electric Flux }

A simple generalization of the Schwinger model on a circle is a theory in 2+1 dimensions with a 3-form field strength $F_{\mu \nu \lambda}$.  The electrically charged object under such a 3-form is an oriented string of charge $e$.  Because a string is a codimension-1 object in 2+1 dimensions, the equations of motion are closely analogous to \eqref{eom}; in particular, they guarantee that the electric flux is constant except when crossing the string, in which case it jumps by $e$ (or $-e$, depending on the orientation and the direction of crossing) \cite{Brown:1988kg}.

Consider such a theory on a spatial cylinder: that is, $x$ non-compact and $y \simeq y+L$.  The initial state is a field strength $F_{\mu \nu \lambda}=Ne \epsilon_{\mu \nu \lambda}$.  For $L\rightarrow \infty$, there is an instanton in this theory which creates a circular loop of string of a critical radius $R_{c}$ (which can be calculated by a method almost identical to that of Section \ref{Dirinst}) and with a nearly identical result\cite{Brown:1988kg}:
\be
R_{c} = {m d \over e(E-e/2)},
\ee
where $d$ is the spacetime dimension of the brane ($d=2$ for a string) and $m$ and $e$ are its tension and charge.

For $R_{c} \ll L$, the  instanton will be essentially unaffected by the periodic boundary conditions, and so for $N>1/2$ the theory is unstable to the nucleation of localized loops of string centered at some point on the cylinder.  The field strength inside the loop is $(N-1)e$; outside it remains $Ne$.  The string is charged under the field, and so  it will expand with constant proper acceleration in both the $x$ and $y$ directions.  After a time of order $L/2$, the string will intersect itself on the opposite side of the circle.  (This scenario is related to those discussed in \cite{Brown:2008ea, Giblin:2010bd}.)

When the string intersects itself, there are again various possibilities.  One is that the string passes through itself without a significant interaction or creation of particles; this is to be expected if the string coupling is small, and probably in the limit of large kinetic energy (although as we will discuss below, the case of D-branes in string theory is more complex).

If the string  passes through itself, the field in the overlap region must have discharged by two units: it is $(N-2)e$.  For $N>1$ there will be a net force on the section of string bounding this region that causes it to continue to expand and accelerate
\begin{figure}
\hspace{-0.5 in}\includegraphics[width=1.2\textwidth]{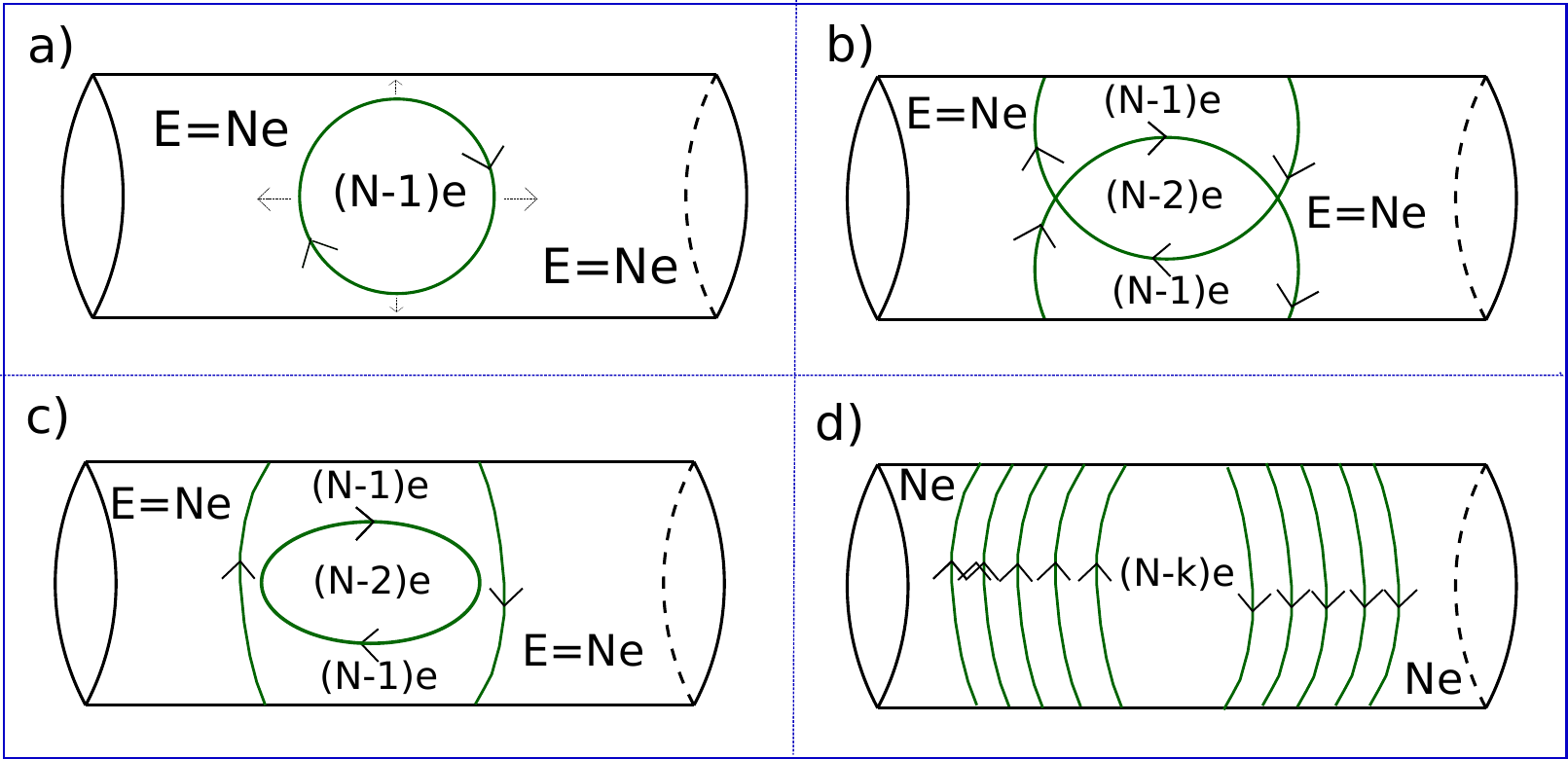}
\caption{\label{cylinder} The evolution of the flux cascade in a $D=2+1$ cylindrical  spacetime with 3-form flux:  a), the string appears and discharges one unit of flux in the region it surrounds.  It expands due to the force exerted by the background field until it intersects itself on the opposite side of the cylinder b), after which  two units of flux are discharged in the intersection region.  String interactions could lead to reconnection at the ``kinks", c), and the system after $k$ wrappings with or without reconnection is schematically d).}
\end{figure}
Eventually this region will collide with itself on the opposite side of the circle, and (if it passes through itself) the field will be discharged by three units in the overlap region, etc.  Absent interactions, due to its kinetic energy in the $y$ direcition the string will continue to wrap around the cylinder even after the flux is fully discharged, and the field at the original nucleation point can oscillate from $Ne$ to $-Ne$ as in the Schwinger model.  However, in the presence of other degrees of freedom, one expects dissipation to damp these oscillations and produce a state with $N \sim 0$ (although precisely if or how this happens is  model dependent).

The simplest string interaction would be to reconnect the string where it crosses itself.  But as can be seen from the figure, this type of interaction does not change the picture substantially, since the reconnection simply produces closed loops of string that will continue to expand and intersect themselves.

From the point of view of the non-compact direction $x$ of the cylinder, this cascading discharge process looks like a series of steps propagating in the $x$ direction, across each of which the flux drops by one unit.  An observer insensitive to dynamics on length scales $L$ and below will not be able to resolve the distance between these steps, and so this will simply look like a 1+1 dimensional ``bubble'' with a wall across which the flux drops by a large number (probably $\sim N$) units.

A curious fact is that the intersection points of the string with itself actually propagate in the non-compact $x$-direction with a speed faster than $c$.\footnote{We thank Ben Freivogel for bringing this  to our attention.}  The same is true for the intersection of points on the string with any line of fixed compact coordinate $y$.  This is possible because such intersection points do not correspond to a fixed part of the string, and it indicates that a naive dimensionally reduced picture of the cascade, where one slices the geometry at fixed $y$ and interprets the string-crossings as massive electrons or positrons in the reduced 1+1 theory, does not behave naturally.  But this simply serves as a warning against such a naive (and in fact incorrect) reduction:  since the geometry is {\it not} independent of $y$, one must retain more than just the Kaluza-Klein zero-modes of the fields to represent the detailed dynamics.  A correct procedure is to average over $y$ or to smear on scales smaller than $L$, in which case the steps in the wall cannot be resolved.  We show in Appendix \ref{AB} in the context of standard electromagnetism on a 3-torus how this kind of reduction can lead to the Schwinger model flux cascade.

\paragraph{Torus with 3-Form Electric Flux }

The dynamics are similar to the scenario above, except that---since the string intersects itself along both spatial directions---the cascade takes place everywhere on the torus. In that sense this model is very close to the Schwinger model on a circle.

\paragraph{Sphere with 3-Form Electric Flux }

Consider a 2+1 theory where the spatial directions form a 2-sphere, in an initial state with a 3-form electric flux.  If the sphere is large compared to $R_{c}$, the standard instanton analysis indicates that there is an instability to nucleate small loops of string at a point on the sphere.  These loops will then expand, reach a maximum size at the equator, and then contract (while continuing to accelerate) onto the antipodal point on the sphere.  The situation is qualitatively different from those considered so far in that the string will now contract to a point in two transverse dimensions.  

In free string theories in flat space, oscillation modes of this type exist.  When the string contracts to a point, since it is non-interactiong it passes through itself rather than ``reflecting''---the next stage of its oscillation is an expansion, {\it but with reversed orientation}.  Since the orientation is reversed, the field the string is charged under must flip sign in the opposite way:  in other words, at least in a spacetime where the string is co-dimension one, the flux will be discharged by two units inside the string after the reversal ({\it c.f.} Figure \ref{cylinder}).

The same conclusion holds if the collapsing string can be represented as a scalar field wall and the kinetic energy in the wall is large enough that self-interactions can be ignored.  In that case, the radial ``free passage'' analysis in Appendix A
 shows that the wall will simply pass through itself in the radial direction---that is, it will go from contracting to expanding, and the field inside it will be discharged by two units.  

If this conclusion is correct, parallel arguments to those made previously indicate that the string will oscillate back and forth between the two poles of the sphere, wiping away an additional unit of flux on each pass and accelerating continuously until the flux is fully discharged---at which point, absent dissipation, annihilation, or other string nucleation events, the flux will begin to increase with the opposite sign.  Whether this happens and how the cascade process terminates is a detailed question that depends on the dynamics of the model.

\subsection{3+1 Dimensional Models}

\paragraph{Cylinder or Torus with 2-Form Electric Flux }

This case is interesting in that the charged objects (point particles) are co-dimension three (greater than one), so they can ``miss'' each other after wrapping around the compact dimension.  We analyze it in some detail in Appendix \ref{AB}.  

Our conclusions are as follows:  
\begin{itemize}

\item When the flux extends in a non-compact direction and both of the other directions form a  torus, when averaged over the torus the model reproduces the results of the Schwinger model on a line (as expected from dimensional reduction). 

\item When the flux wraps a single compact dimension and the other two are non-compact, the pair  circles the compact direction indefinitely,  affecting the background field  slightly inside the future lightcone of the nucleation point.

\item  When all three directions are compact, the model reproduces our result for the Schwinger model on a circle ({\it i.e.}, there is a flux cascade).  The time dependence of the average $E$-field is
\be
\langle E(t) \rangle \sim E_{0}  - {8 \pi e t \over L_{x} L_{y} L_z},
 \ee
where $L_i$ is the length of the circle in the $i$-direction.  
\end{itemize}
We expect analogous results to hold for 2-forms  ({\it i.e.} point particles) in any number of dimensions.

\paragraph{One Compact Direction with 3-Form Electric Flux } \label{fountain}

For definiteness, suppose $x \simeq x + L$, $y,z$ are non-compact, and the initial flux is $F_{txy}=E_{0}$.  Assuming as usual that $R_{c} \ll L$, a circular string extending in the $x,y$ plane will appear centered at some point---which we choose to be $x=y=z=0$---and then stretch in the $x,y$ plane.  After a time of order $L/2$ it will intersect itself as it wraps the $x$-direction.  In the non-interacting limit, it will continue to grow and intersect itself in the $x,y$ plane.

Because the $z$ direction is non-compact this does not lead to a flux cascade even near the nucleation point $\vec x = 0$.  From the point of view of the dimensionally reduced (on $x$) theory, this process looks surprising---the naive picture is of a ``fountain'' of electrons and positrons spewing out from $y=z=0$, with the line of positrons accelerating in the $+y$ directions and the line of electrons accelerating in the $-y$ direction.  However the spacing between the particles is ${\cal O}(L)$, and so (as mentioned above) one is not really justified in regarding them as individual particles in the reduced theory.  Nevertheless, the process would look rather exotic to a low-energy observer.

\subsection{$4+1$ Dimensional Models}
\paragraph{ One Compact Dimension, with 4-Form Electric Flux }
Suppose $w \simeq w+L$, $x,y,z$ are non-compact, and the initial flux is $F_{txyw}=E_{0}$.
This  is a higher dimensional version of the scenario discussed in \ref{fountain}.  If the flux wraps the compact cycle and two of the three non-compact directions, the nucleated object is a 2-sphere that appears at a point in $z$ and expands in the $x,y$ plane, while intersecting itself in the $w$ direction.  From the dimensionally reduced point of view this is a series of  expanding circular strings concentric with the nucleation point, not unlike ripples on a pond (although again, a low-energy observer cannot resolve the individual ripples).  Due to the non-compact direction $z$ there is no flux cascade.

\subsection{String and M-Theory}

Naively, the cases analyzed above generalize in a simple way to discharge of ``electric'' flux in string or M-theory.  However, we have used a weak interaction approximation to understand the evolution of the charged objects after they appear and as they intersect themselves.  The charged objects of string theory---D-branes---are not really weakly coupled even in the limit of small string coupling $g_{s}\rightarrow 0$.

For instance, one issue specific to the case of colliding D-branes in string theory is the production of open strings stretched between the branes.  Because the mass of these strings depends on the brane separation, and because of the exponentially growing density of such open string states, this can make brane-brane collisions extremely inelastic at ultrarelativistic velocities even for very small values of $g_{s}$ \cite{Bachas:1995kx, McAllister:2004gd}.

Our case differs from that of parallel brane-brane collisions in two important ways.  Locally, collisions between a spherical brane and itself as it wraps a compact direction are more akin to brane-anti-brane collisions than to brane-brane collisions.  In particular, one expects a tachyon in the spectrum of open string states, at least in regions where the brane intersects or comes within a string length of itself.   The other difference is that our branes are curved and undergoing constant proper acceleration, which affects the dynamics of string creation and tachyon condensation.

The tachyonic stretched string between parallel stationary branes and anti-branes condenses in a time of order $1/m_{s}=l_{s}$, leading to the annihilation of the branes into closed string modes \cite{Sen:1998sm, Kleban:2000pf}.  However in our case the collision is relativistic and there is very little time for this condensation to occur.  The lowest stretched string mode is tachyonic only for separations equal to or less than $l_{s}$, a situation that lasts only for a time of order $l_{s}$.  Moreover, in the center of mass frame of the collision, all processes---including tachyon condensation---should be time dilated and hence slowed by a factor of $1/\gamma$.  

For these reasons it seems unlikely, at least in the case of relativistic velocities, that tachyon condensation will generically lead to the annihilation between large sections of the brane and anti-brane (or oppositely oriented piece of brane).  We study this question in detail in Appendix \ref{AC} by computing the effective action for the would-be tachyon and solving its equations of motion.  The result is in fact that the ``tachyon'' does not condense significantly when $R_{c} \ll L$ and the collision is relativistic.

The force from open strings created between colliding parallel branes is independent of distance (at least in the case of small impact parameter), because the mass of the strings grows linearly with the brane-brane separation.  As such, it has the same distance dependence ({\it i.e.} none) as the force from an external field.  The tension between the two branes exerted by quantum mechanical production of open strings is roughly $\epsilon \gamma (\ln \gamma)^{p}$ in string units, where $\epsilon$ is a small dimensionless prefactor ($\approx 1/3500$ for the case of 5-branes), and $p$ is an ${\cal O}(1)$ number \cite{McAllister:2004gd}, while the force exerted by $N$ units of flux is $\sim N$ in these units.  Therefore even in this case, the force from created strings can overwhelm the force from the field only if the velocity is such that $\gamma \sim N$.  
Of course, once all the flux has been discharged, even a weak force from string creation plus tachyon condensation could lead to the branes annihilating into closed string radiation and prevent the flux from re-charging with the opposite sign.

From this heuristic analysis it appears that stretched string creation is simply a source of friction rather than a tether that completely halts the cascade.  One expects that it may help bring the flux to rest near zero.  Clearly, a more complete analysis in a specific model would be needed to work out the detailed dynamics.

\subsection{Bousso-Polchinski Model}

In the  Bousso-Polchinski \cite{discretuum} toy model for the string landscape, contributions to the 4D vacuum energy come from the 7-form field strength $F_{7}$ of 10+1D M-theory, with three legs of the $F_{7}$ wrapped on some compact three cycle $C_{i}$, and the other four extended in 4D spacetime.  One unit of this 7-form electric flux can be discharged by the nucleation of a spherical M5 brane of radius $R_{c}$.  So long as $R_{c} \ll L_{i}$, where $L_{i}$ is the length scale associated with the 3-cycle $C_{i}$, the brane will appear as a 5-sphere extended in the 3 non-compact space dimensions and wrapping the three dimensions of $C_{i}$.  The sphere will expand and accelerate after it nucleates, wrap around $C_{i}$, and intersect itself.  

One might worry that the radius $R_{c}$ of the instanton might for some reason always be larger than $L_{i}$.  However it is simple to see that this is not the case.   Suppose we can reduce on the  four compact dimensions without flux, which have volume $L^{4}$.  The theory becomes a 6+1 dimensional theory, and the 5-brane is codimension one.  The critical radius of the bubble is given by the following formula:
\be \label{crit11}
R_{c} = {\tau \over e_{7} E_{7}} = {2 \pi M_{11}^{6} \over N e_{7}^{2}} = {M_{11}^{3} L^{4} \over 2 \pi N},
\ee
where $\tau = 2 \pi M_{11}^{6}$ is the tension of the M-5 brane, $e_{7} = 2 \pi M_{11}^{3/2}/L^{2}$ is the effective 7-dimensional charge in the reduced theory, and $N$ is the number of units of flux.  %CHECK FACTORS OF 2 AND PI 

The 7-dimensional energy density in $N$ units of flux is $\rho_{N} \sim (N e_{7})^{2} = 2 \pi^{2} N^{2} M_{11}^{3}/L^{4}$.  If we require that the energy density in the flux be less than the Planck energy density in 7D $\rho_{Pl}\sim M_{11}^{11}L^{4}$, there is a maximum possible $N$:
\be \label{Nmax}
 2 \pi^{2} N^{2} M_{11}^{3}/L^{4} < M_{11}^{11}L^{4} \longrightarrow N < \left( {M_{11} L} \right)^{4} {1 \over 2 \pi}.
 \ee
 Putting this into \eqref{crit11} gives simply
 \be
 R_{c} > 1/M_{11}.
\ee 
(This result also follows from dimensional analysis:  there are no dimensionless parameters in the 11D theory, and the dynamics here are independent of the volume of the 4-manifold $L^{4}$.)  

The remaining three compact dimensions can have any length $L_{3} > 1/M_{11}$.  Therefore, there seems to be no reason why the would-be critical bubble must have a radius larger than the size of the cycle it expands into, and---at least for weak coupling and neglecting back-reaction---when $R_{c} < L_{3}$ the 5-brane will intersect itself and will be able to discharge the flux near its center in a flux cascade.  As mentioned above, from the 4D point of view this looks like an expanding bubble inside of which the flux has been discharged by many units.

Of course, there are many caveats.  If the 3-cycle the flux wraps is non-trivially fibered over the 4-manifold, one cannot reduce in the naive way we have done.  We have neglected back-reaction, which will certainly become important if the cascade discharges of order the maximum number of units of flux allowed by \eqref{Nmax}.  We have not taken radiation of gravitons or production of light stretched M-2 branes into account, and we have only considered a single type of flux.

\subsection{Brane-Flux Annihilation, etc.}

Analyzing the dynamics of flux cascades in  explicit string compactifications goes beyond the scope of this work.  Generally speaking, any string vacuum with net positive vacuum energy is at best metastable, and one of the typical decay channels is discharge of  flux by the nucleation of a brane.  it does not seem to require fine-tuning for the critical radius of the nucleated brane to be somewhat smaller than the size of the compact dimensions, and then, as we have seen, the flux cascade process is generic at least at weak coupling.  It would be interesting to study the cascade in scenaria such as that of \cite{Kachru:2002gs} or \cite{kklt}.

We expect flux cascades to have a significant effect on analyses of the dynamics of transitions and probability measures on vacua in the string landscape.  As one example, the ``straggering'' phenomenon of \cite{SchwartzPerlov:2006hi} could be strongly affected.

\section*{Acknowledgements}
We thank Jose Blanco-Pillado, Adam Brown, Willy Fischler, Ben Freivogel, Lam Hui, Jaume Garriga, Steve Shenker, Leonard Susskind, and I-Sheng Yang for discussions.  This material is based upon work supported in part by the National Science Foundation under Grant No. 1066293 and the hospitality of the Aspen Center for Physics.  
The work of MK is supported by NSF CAREER grant PHY-0645435. M.P. is supported in part by NSF grant PHY-0758032, and by ERC Advanced Investigator Grant n.226455 {\em Supersymmetry,
Quantum Gravity and Gauge Fields (Superfields)}.

\appendix

\section{Spherical Free Passage} \label{AA}

In this Appendix we will demonstrate that the ``free passage'' approximation of \cite{Giblin:2010bd} extends to the case of spherically symmetric collapsing regions in arbitrary spacetime dimensions.

In Lorentz invariant models in which bubbles of a lower energy state can form---for instance scalar field theories with multiple meta-stable minima with different vacuum energy, or the models with fluxes and electrically charged objects considered in this paper---the walls of the bubble typically accelerate with constant proper acceleration.  As a result, after a few characteristic times the walls are ultra-relativistic, and therefore the collision between two such walls tends to be dominated by kinetic energy, rather than by interactions.  For the case of scalar field theories, this presumably means that the collision can be treated using free field theory, at least for times short compared to the timescale set by the interactions (such as the height of the relevant potential energy barriers).

In \cite{Giblin:2010bd} it was argued that for the case of colliding parallel planar free field walls that separate a middle region with $\phi=\phi_{F}$ from regions with $\phi=\phi_{L}$ and $\phi=\phi_{R}$, the field value after the collision in the middle  region is $\phi = \phi_{F} + (\phi_{L}-\phi_{F})+(\phi_{R}-\phi_{F})$.  Here, we extend the argument to collapsing spherical field configurations in arbitrary dimensions, where the field at the center is initially $\phi=\phi_{F}$.\footnote{MK would like to thank L. Hui, I. Yang, and B. Freivogel for discussions on this point.}  Since for spherical symmetry the field at large radius satisfies $\phi_{L}=\phi_{R}\equiv \phi_{T}$, the ``free passage'' expectation is that the field at the center is $\phi = 2 \phi_{T}-\phi_{F}$ for times after the collapse.

To be precise, we will consider free scalar field theory in $n+1$ spacetime dimensions, where the initial configuration $\phi(\vec x, t=0) \equiv \phi_{0}(\vec x)$  is spherically symmetric and satisfies 
\bea \label{initphi}
 \phi_{0} = \phi_F ~(r \ll R), ~~ ~~\phi_{0}  = \phi_{T}~ (r \gg R),
\eea
For the moment we will leave $\dot \phi_{0}$ unspecified, but we are only interested in initial conditions representing collapsing bubbles, {\it i.e.} those for which 
\bea \label{limits}
\phi(r =0, t\rightarrow -\infty) =\phi_F,  \\
\phi(r\rightarrow \infty, t) = \phi_{T}.   
\eea
In any number of dimensions, the value of a free scalar field at an arbitrary spacetime point in $n+1$ dimensions is determined by its initial data at $t=0$ by a formula of the form
\bea \label{phixt} \nonumber
\phi(\vec x, t) = C_{n} \partial_{t^{2}}^{ n-3 \over 2}  t^{n-2} Q[\dot \phi_{0}] +  \partial_{t} \left(  C_{n} \partial_{t^{2}}^{(n-3)/2}  t^{n-2} Q[ \phi_{0}] \right) ~~(n~{\rm odd}) \\ 
\phi(\vec x, t) = C_{n} \partial_{t^{2}}^{n-2 \over 2}  t^{n-3} H[\dot \phi_{0}] +  \partial_{t} \left[   C_{n} \partial_{t^{2}}^{(n-2/2}  t^{n-3} H[\phi_{0}]\right] ~~(n~{\rm even}).
\eea
Here $\phi_{0} = \phi(\vec x,0), \dot\phi_{0}=\dot\phi(\vec{x},t)|_{t=0}$ is the initial data, $C_{n}$ is a constant, and $Q(\phi_{0}) = \omega_{n}^{-1} \int^{|\vec \beta| = 1} \phi(\vec x + \vec \beta t) d \vec \beta$ and $H(\phi_{0}) =  \int^{t} Q(\vec x, r) r (t^2-r^2)^{-1/2} dr$ are integrals are carried out over a spherical shell or volume of radius $t$ centered on $\vec x$ (see \cite{ch}, p. 682 for  explicit expressions).  

Focussing on $\vec x = 0$, schematically \eqref{phixt} takes the form
\be
\phi(0, t) = A_{1}[\dot \phi_{0}](t) + A_{2}[\phi_{0}](t),
\ee
where $A_{1}$ and $A_{2}$ are integrals over balls or spheres of radius $|t|$ centered on $\vec x = 0$.

The key to our argument is that $A_{1}$ is an odd function of $t$ (with the initial data $\dot \phi_{0}$ held fixed), while $A_{2}$ is an even function of $t$.  As $t \rightarrow -\infty$, it is simple to see that the initial data \eqref{initphi} means that $A_{2} \rightarrow \phi_{T}$.  Therefore  \eqref{limits} requires a choice of $\dot \phi_{0}$ such that $A_{1} \rightarrow \phi_F-\phi_{T}$.  But this implies by $t \leftrightarrow -t$ symmetry that as $t \rightarrow +\infty$, $\phi(0, t) = A_{1} + A_{2} \rightarrow 2 \phi_{T}-\phi_F$, which is the naive spherical analog of the ``free passage'' result.

\section{Schwinger Pair Production on a Torus} \label{AB}

In this Appendix we consider the problem of standard (3+1 dimensional) Schwinger pair production, but where some or all of the spatial dimensions are periodically identified (compactified on a torus).  We  expect the generalization to other numbers of dimensions to be straightforward.  

There are two representative cases: (1) the initial state $E$ field is oriented along a non-compact direction and the transverse dimensions are compact, and (2) the $E$ field is oriented along a compact direction with the two transverse dimensions non-compact.

To fix notation, the initial state field $\vec E_0 = E_0 \hat x$, and we will use lightcone coordinates $u=(t+x)/\sqrt{2}$ and $v=(t-x)/\sqrt{2}$, with coordinates $\rho, \phi$ or $y,z$ parametrizing the two directions transverse to $x$.  We will work in the approximation that the charges move at speed $c$ from the moment they appear.  In addition we will ignore scattering (which could be relevant in cases where the $E$ field wraps a compact dimension).

 Before imposing the periodic boundary conditions, the electromagnetic field due to a single electron-positron pair created from the vacuum at $x=t=\rho=0$, which then move apart relativistically in the $\pm x$-direction, is
\bea \label{EM}
\vec E_p = {-4 e \over \rho} \delta_+(\vec x^2)\left( \rho \hat x - x \hat \rho \right) \\ \nonumber
\vec B_p = {-4 e} \delta_+(\vec x^2) {\sqrt{\rho^2 + x^2} \over \rho} \hat \phi,
\eea
where $\delta_+(\vec x^2) \equiv \theta(u)\theta(v)\delta(2uv - \rho^2)$ is a delta function on the surface of the future lightcone of the point $x=\rho=t=0$.  (We will not derive these fields here; the derivation can be found in \cite{Aichelburg:2003fe}, c.f. Eq. (8).  
Note that this is {\it not} the field of a single relativistic charge moving with constant velocity, which has a non-zero field strength at all times in the past---it is the field of a pair created at $t=0$ and has non-zero field strength only for $t>0$.)
The most salient feature of this field configuration is that it is confined the surface of the future lightcone of the nucleation point:  that is, the fields are non-zero only on the surface of a sphere that expands at speed $c$ from $x=\rho=t=0$.

\paragraph{Case (1):} Take $x$ non-compact, $y \simeq y+L_y$ and $z\simeq z+L_z$.   Imposing these conditions requires summing the fields \eqref{EM} over image points situated at $y=n L_y, z=m L_z$, for all integer pairs $(n,m)$.   The result is a rectangular lattice (in the $y,z$ plane) of  expanding spherical shells.  If we reduce on the $y,z$ directions to produce a 1+1 dimensional theory, we should obtain a result consistent with the Schwinger model: that is, the field of the charged pair should be $E_p=0$ for $x \gg t$, and $E_p=e_2$ for $x \ll t$, where $e_2$ is the effective  charge in the 2D reduced theory.

To see this, we should reduce on the $y,z$ directions by averaging over a cell of area $L_y L_z$ of the $yz$ lattice.  Because the lattice is invariant under $y \rightarrow -y$ and $z \rightarrow -z$, all field components in the $y, z$ directions necessarily average to zero.  Therefore from \eqref{EM}, $\langle \vec B_p \rangle = 0$, and  $\langle \vec E_p \rangle \propto \hat x$.  To compute $\langle \vec E_p \rangle $, note that the integral of the $x$-component of $\vec E_p$ over the $y,z$ plane at fixed $u,v>0$ is simply
\be
\int 2 \pi \rho d\rho (-4 e) \delta(-t^2 + x^2 + \rho^2) = -4 \pi e.
\ee
Since the density of lattice points in the $yz$ plane is $1/(L_y L_z)$, we obtain
\bea \nonumber
\langle \vec E_p \rangle  &=& -{4 \pi e \over L_y L_z}, ~~  x \ll t \\ \nonumber
\langle \vec E_p \rangle  &=& 0,~~  x > t,
\eea
which demonstrates that the averaged field reproduces the 1+1D result.  

This also nicely illustrates our main result in another way:  if $x \simeq x+ L_x$ is compactied along with $y$ and $z$, the field due to the single charged pair will discharge the initial state field in a cascade as more and more  ``shells" overlap due to the image pairs along the $x$-axis:
\be
\langle \vec E_p \rangle  \sim -{4 \pi e \over L_y L_z} {2 t \over L_x} \\ \nonumber
\ee

 \paragraph{Case (2):}  Here $x \simeq x+L$, and $y,z$ are non-compact.  In the approximation that we can ignore scattering, the total field is a sum over images in the $x$ direction.  The result is an infinite set of expanding spheres centered at regular intervals $x=n L$ along the line $\rho=0$.  A glance at \eqref{EM} shows that the fields are zero for $\rho > t$.  For fixed $\rho$ as a function of $t$, it is clear that the fields become periodic with frequency $c/L$ at late times.  This follows because for $t \gg \rho$ the expanding spheres are very large, so that successive wave-fronts are nearly parallel, and by the same token each successive shell arrives with closer and closer to equal time spacing $L/c$.  Therefore at fixed $\rho$, the time average of the fields over  time intervals $\sim L/c$   converges to a constant.
 
From this we see that the effect on the background field is to modify the fields slightly and by a constant (on average) amount everywhere in the region $\rho < t$.  The charges will continue circling the $x$-direction indefinitely so long as $y$ and $z$ are truly non-compact and no other pair creation events occur.

\section{Tachyon Condensation During Wrapped Brane Self-Collisions} \label{AC}

To analyze the dynamics of the open string tachyon, it is convenient to introduce Milne-type coordinates as in \cite{Brown:2008ea}.\footnote{We thank A. Brown for suggesting this approach.}  In these coordinates the collision of an expanding spherical brane with one of its covering space images happens at one instant of time, everywhere along a spatial hyperboloid.  In other words, apart from a warp factor due to the curvilinear nature of the coordinates, the self-collision is not very different from the collision between parallel flat branes and anti-branes that approach each other.

For simplicity we consider a 2+1 dimensional model on a cylinder with ``D1 branes" charged under a 3-form field strength.  The generalization to higher dimensions is straight forward.  Cartesian coordinates on the spacetime are $x\sim x+L $, $y$ and $t$. The relevant Milne-type coordinates are:
$$
t + y = e^{U}, \qquad t - y = e^V, \qquad U + V = 2\tau, \qquad U - V = 2\zeta,
$$
and the metric in these coordinates is:
$$
ds^2 = e^{2\tau} (-d \tau^2 + d\zeta^2) + dx^2.
$$

The D1 brane accelerating under the influence of the background electric flux is described by the equation $x^2 = e^{2\tau} + R_c^2$, with $R_c$ the initial nucleation
radius. The induced metric of a fundamental string stretching between a point on the D1
and one of the mirror images is determined by giving  $\zeta$ vs. $x, \tau$: $  \zeta= \zeta (x,\tau )$. The induced
metric on the 2d world sheet of the string is thus
$$
ds^2_{string} = e^{2\tau} (-1 + \dot{\zeta}^2)d \tau^2 + (e^{2\tau} \zeta'^2 + 1)dx^2.
$$
The action is
$$
S = m_s^2 \int dxd \tau e^\tau \sqrt{(1 -\dot{\zeta}^2)(e^{2\tau} \zeta'^2 + 1)}
$$
One obvious solution is $\zeta = constant$, i.e. a static string stretching from a point
$x = +\sqrt{e^{2\tau} + R_c^2}$ to the point $x = -\sqrt{e^{2\tau} + R_c^2}$.
The energy of this string follows from the action (notice the warp factor):
$$
E = m_s e ^\tau  \min_{n\in Z}| 2\sqrt{e^{2\tau} + R_c^2}-nL|.
$$

Define new time $T=e^\tau$, then energy is 
\be \label{ET}
E_T= m_s \min_{n\in Z}| 2\sqrt{T^2 + R_c^2}-nL|.
\ee
The metric on the D1  world sheet $x=\pm\sqrt{T^2 + r^2}$ is 
$$
ds_{D1}^2= T^2 d\zeta^2 - (T^2 + R_c^2)^{-1}R_c^2dT^2,
$$
which asymptotes to the 2d de Sitter metric $ds^2_{D1}= -R_c^2 T^{-2}dT^2 + T^2 d\zeta^2$ for $T \gg R_c$.  Therefore the action of the open string tachyon $\Phi$ is
\be \label{tachac}
S=\int d\tau d\zeta (R_c/2) [ -(T/R_c)^2(d\Phi/dT)^2 + T^{-2} (d\Phi/d\zeta)^2 + (E_T^2-m_{String}^2)\Phi^2].
\ee
As can be seen, the ``tachyon" $\Phi$ becomes truly tachyonic for the range of times where $E_T < m_s$; in other words, when the brane is within a string length $l_s=1/m_s$ of one of its images.

Clearly, if $\Phi$ remains tachyonic for all times, it would condense and lead to the annihilation of the brane and anti-brane.  However, for relativistic collisions $\Phi$ is tachyonic only for a time of order $l_s$.  Moreover, as we will see the curvature and acceleration of the string plays a role.  The equations of motion that follow from \ref{tachac} are
\be \label{tacheq}
(T/R_c)^2 \ddot \Phi + 2 (T/R_c^2) \dot \Phi -T^{-2} \partial^2 \Phi/d \zeta^2 - (E_T^2 - m_s^2)\Phi =0.
\ee
Assuming $\Phi=\Phi(T)$ depends on time only, one can solve these equations  in terms of Whittaker functions.  However, a simple approximation suffices to illustrate the relevant physics.  The field $\Phi$ can only condense when its effective mass term is negative.  Therefore, to be as conservative as possible consider $T \approx nL/2$ when the mass term reaches its largest negative value, and ignore the ``friction" term $ \dot \Phi$.  In this approximation \eqref{tacheq} reduces to
$$
\ddot \Phi = - \left( 2 m_s R_c/ n L \right)^2 \Phi \rightarrow \Phi \sim e^{2 m_s R_c T/ n L}.
$$
From \eqref{ET} the tachyonic phase lasts for a time $\Delta T \approx l_s$.  Therefore $\Phi$ can only grows at most by an amount $\Delta \Phi/\Phi \sim 2 R_c/ n L\ll 1$ for $R_c \ll L$ at each collision.  For times in between the collisions $\Phi$ oscillates and damps (due to the friction term).  Therefore we do not expect that brane self-collisions in this regime of parameters can lead to annihilation.  

A very similar analysis applies to higher dimensional co-dimension one branes.  Branes with  co-dimension larger than one may have a non-zero impact parameter, in which case the lightest open string mode may never become tachyonic, and in all cases has a more positive mass than in the case of coincident brane-anti-brane pairs.

\bibliographystyle{klebphys2}

\bibliography{schwing}

\end{document}